\documentclass[draft,grl]{agu2001}
\authorrunninghead{ESELEVICH ET AL.}

\titlerunninghead{COMMENT}

\authoraddr{V. G. Eselevich, Institute of Solar-Terrestrial Physics, Siberian Branch of Russian
Academy of Sciences, Lermontova 126a, P.O. Box 291, Irkutsk
664033, Russia. (esel@iszf.irk.ru)}

\authoraddr{V. M. Bogod, Special Astronomical (Pulkovo) Observatory,
Russian Academy of Sciences, Pulkovskoe sh. 65, St. Petersburg,
196140 Russia. (vbog@sao.ru, vbog@gao.spb.ru)}

\authoraddr{I. V. Chashei, Pushchino Radio Astronomy Observatory, Lebedev
Physics Institute, Russian Academy of Sciences, Pushchino, 142292
Russia. (chashey@prao.ru)}

\authoraddr{M. V. Eselevich, Institute of Solar-Terrestrial Physics, Siberian Branch of Russian
Academy of Sciences, Lermontova 126a, P.O. Box 291, Irkutsk
664033, Russia. (mesel@iszf.irk.ru)}

\authoraddr{Yu. I. Yermolaev, Space Research Institute, Russian Academy of
Sciences, Profsoyuznaya ul. 84/32, Moscow, 117997 Russia.
(yermol@iki.rssi.ru)}

\begin{document}

%
%
%
%
%

%
%

\title{Comment on ``CAWSES November 7-8, 2004, superstorm: Complex solar and
interplanetary features in the post-solar maximum phase'' by B. T.
Tsurutani, E. Echer, F. L. Guarnieri, and J. U. Kozyra}

%
%






\authors{V. G. Eselevich\altaffilmark{1},
V. M. Bogod\altaffilmark{2}, I. V. Chashei\altaffilmark{3}, M. V.
Eselevich\altaffilmark{1}, and Yu. I. Yermolaev\altaffilmark{4}}

\altaffiltext{1} {Institute of solar-terrestrial physics,
Siberiain Branch of Russian Academy of Sciences, Irkutsk, Russia.}

\altaffiltext{2} {Special Astronomical (Pulkovo) Observatory,
Russian Academy of Sciences, St. Petersburg, Russia.}

\altaffiltext{3} {Pushchino Radio Astronomy Observatory, Lebedev
Physics Institute, Russian Academy of Sciences, Pushchino,
Russia.}

\altaffiltext{4} {Space Research Institute, Russian Academy of
Sciences, Moscow, Russia.}


%
%
%



%
%


\begin{abstract}
(No abstract for comment)
\end{abstract}

%
%

%

\begin{article}

%
%

Recently \citet{Tsurutani2008} (Paper 1) analyzed the complex
interplanetary structures during 7 to 8 November, 2004 to identify
their properties as well as resultant geomagnetic effects and the
solar origins. Besides mentioned paper by \citet{Gopalswamy2006}
the solar and interplanetary sources of geomagnetic storm on 7-10
November, 2004 have also been discussed in details in series of
papers \citep{Ishkov2005,Yermolaev2005,Wintoft2005,Chertok2006,
Arkhangelskaja2006,Trichtchenko2007,Culhane2007}. Some conclusions
of these works essentially differ from conclusions of the Paper 1
but have not been discussed by authors of Paper 1. In this comment
we would like to discuss some of these distinctions.

\citet{Tsurutani2008} studied 3 fast interplanetary shocks (marked
by FS1, FS2 and FS3) observed on 7 November before passage of
magnetic cloud resulting in strong magnetic storm and indicated
correspondences between these shocks and possible flares. These
relations are summarized in Table 1 where 1st column shows shock
number, 2nd -- time of shock arrival, 3rd presents data for 2
corresponding flares for each shock -- X-ray importance, time and
date, active region (coordinates) on the Sun. Next 2 columns
present data about flares (the same data for flares accompanied by
corresponding CMEs) and CMEs (time and date occurrence in C2
coronagraph of SOHO/LASCO instrument, type of CME, velocity in
field of view) resulting in these shocks obtained by
\citep{Yermolaev2005} (Paper 2). It is important to note that in
Paper 1 the sources of shocks are suggested to be flares while in
Paper 2 -- CMEs. Thus calculated times of flight from the Sun up
to the Earth for FS1 and FS2 shocks in Paper 1 are about 1 day
more than in Paper 2.

We think that the reasons of these distinctions are the following:
\begin{enumerate}
\item The empirical relations which have been checked up on the big
experimental material \citep{Cane2003,Eselevich2004} were used for
calculation of time of flight in Paper 2, and also some features
of their application \citep{Cane1986a,Sheeley1985a,Sheeley1985b}
were taken into account. Recent papers on the same problem
\citep{Gopalswamy2007,Kim2007} give similar estimations of time of
flight for these shocks. Method of calculation of flight time in
details is not described in Paper 1, but it is noted that velocity
of a shock is considered to be constant on the Sun -- Earth way.
This assumption cannot be executed for FS1 and FS2 as velocity of
CMEs connected with corresponding flares in two and more times
exceeds the maximal velocity of plasma $V_{max}$ behind shock
fronts near the Earth.
\item As in Paper 1 the flares are supposed to be sources
of shocks, authors of Paper 1 suggested that blast shock can reach
an orbit of the Earth. However till now there is no any
experimental proof, that blast shock can reach the Earth. At the
same time practically for all CMEs in corona which move in the
Earth direction and have speed more than 400 km s$^{-1}$ in
corona, the shock in an orbit of the Earth is registered
\citep{Sheeley1985a,Sheeley1985b,Cane1986b,Eselevich2006}.
Therefore it has been naturally assumed in Paper 2 that CMEs are
sources of sporadic streams of solar wind near the Earth.
\end{enumerate}

Thus, conclusions of Paper 1 about sources of 3 shocks and times
of their motion differ from results of early published papers and
in our opinion are incorrect.

\begin{acknowledgements}
       Work was in part supported by RFBR, grant 07-02-00042.
\end{acknowledgements}

\end{article}

\clearpage

\begin{table}
\caption{Solar events, resulting in fast interplanetary shocks on
7 November, 2004}
\begin{flushleft}
\begin{tabular}{lcllll}
\tableline \multicolumn{1}{c}{Shock}&Time,&\multicolumn{1}{c}{{\it
Tsurutani et al.}, 2008}&\hskip1pt&\multicolumn{2}{c}{{\it
Yermolaev et al.}, 2005}
\\ \cline{3-3}\cline{5-6} \noalign{\vskip4pt}
 &UT&Flare&&Flare&CME
\\ \tableline
FS1&01:55&C1.8 at 01:22 UT Nov. 2&&M4.7 at 15:24 UT Nov. 3&15:54
UT Nov. 3
\\ &&AR 693 (S17E10) or&&(N04E37)&partial halo CME
\\ &&C6.9 at 01:43 UT Nov. 2&&&V$_k$ = 800 km s$^{-1}$
\\ &&AR 687 (N11W92)&&&
\\ \tableline
FS2&10:00&M2.8 at 01:28 UT Nov. 3&&C6.0 at 09:00 UT Nov. 4&09:54
UT Nov. 4
\\ &&AR 691 or&&(N03E27)&full halo CME
\\ &&M1.6 at 03:32 UT Nov. 3&&&V$_k$ = 550 km s$^{-1}$
\\ &&AR 696 (N09E45)&&&
\\ \tableline
FS3&17:55&M2.5 at 23:00 UT Nov. 5&&M5.5 at 21:42 UT Nov. 4&00:30
UT Nov. 5
\\ &&AR 696 (N11E19) or&&(N05E18)&full halo CME
\\ &&M5.4 at 23:09 UT Nov. 5.&&&V$_k$ = 720-1100 km s$^{-1}$
\\ &&AR 696&&&
\\ \tableline
\end{tabular}
\end{flushleft}
\end{table}


\begin{thebibliography}{}

  \bibitem[{\it Arkhangelskaja et al.}(2006)]{Arkhangelskaja2006}
\reference
Arkhangelskaja, I. V., A. I. Arkhangelsky, Yu. D.
Kotov, S. N. Kuznetsov, and A. S. Glyanenko (2006), The Solar
Flare Catalog in the Low-Energy Gamma-Ray Range Based on the AVS-F
Instrument Data Onboard the CORONAS-F Satellite in 2001-2005, {\it
Solar System Research}, {\it 40}(2), 133-141.

  \bibitem[{\it Cane et al.}(1986a)]{Cane1986a}
\reference
Cane, H. V., R. E. McGuire, and T. T. Von Rosenvince
(1986), Two classes of solar energetic particle events associated
with impulsive and long-duration soft X-ray flares, {\it
Astrophys. J}, {\it 301} 449.

  \bibitem[{\it Cane et al.}(1986b)]{Cane1986b}
\reference
Cane, H. V., S. W. Kahler, and Sheeley, N. R. Jr.
(1986), Interplanetary shock preceded by solar filament eruptions,
{\it J. Geophys. Res.}, {\it 91}, 13321.

  \bibitem[{\it Cane and Richardson}(2003)]{Cane2003}
\reference
Cane, H. V., and I. G. Richardson (2003),
Interplanetary coronal mass ejections in the near-Earth solar wind
during 1996-2002 , {\it J. Geophys. Res.}, {\it 108}(A4), SSH 6-1.

  \bibitem[{\it Chertok}(2006)]{Chertok2006}
\reference
Chertok, I. M. (2006), Large-Scale Activity in Major
Solar Eruptive Events of November 2004 According to SOHO Data,
{\it Astronomy Reports}, {\it 50}(1), 68-78. Translated from
Astronomicheski Zhurnal, 2006, Vol. 83, No. 1, pp. 76-87.

  \bibitem[{\it Culhane et al.}(2007)]{Culhane2007}
\reference
Culhane, J.L., S. Pohjolainen, L. van Driel-Gesztelyi,
P. K. Manoharan, and H. A. Elliott (2007), Study of CME transit
speeds for the event of 07-NOV-2004, {\it Advances in Space
Research}, {\it 40}(12), 1807-1814, doi:10.1016/j.asr.2007.01.005.

  \bibitem[{\it Eselevich and Eselevich}(2004)]{Eselevich2004}
\reference
Eselevich, M. V., and V. G. Eselevich (2004), Sporadic
plasma streams and their sources in the period of extraordinary
solar activity from October 26 to November 6, 2003, {\it Cosmic
Research}, {\it 42}(6), 571-582.


  \bibitem[{\it Eselevich and Khlystova}(2006)]{Eselevich2006}
\reference Eselevich, M.V., and A.I. Khlystova (2006),
Relationship between the 195 \AA\  flare parameters and the halo
CME velocities, 9th Young Scientists' Conference ``Physical
Processes in Space and Near-Earth Environment'', 11-16 September
2006, Irkutsk, Russia, p. 192 (in Russian).

  \bibitem[{\it Gopalswamy et al.}(2006)]{Gopalswamy2006}
\reference
Gopalswamy, N., S. Yashiro, and S. Akiyama (2006),
Coronal mass ejections and space weather due to extreme events,
Proceedings of the ILWS Workshop. Goa, India. February 19-24,
2006. Editors: N. Gopalswamy and A. Bhattacharyya. ISBN:
81-87099-40-2, p. 79, 2006.

  \bibitem[{\it Gopalswamy et al.}(2007)]{Gopalswamy2007}
\reference
Gopalswamy, N., S. Yashiro, and S. Akiyama (2007),
Geoeffectiveness of halo coronal mass ejections, {\it J. Geophys.
Res.}, {\it 112}, A006112, doi: 10.1029/2006JA012149, 2007.

  \bibitem[{\it Ishkov}(2005)]{Ishkov2005}
\reference Ishkov, V. N., Properties of the Current 23rd
Solar-Activity Cycle (2005), {\it Solar System Research}, {\it
39}(6), 453-461. Translated from Astronomicheskii Vestnik, Vol.
39, No. 6, 2005, pp. 507-516.

  \bibitem[{\it Kim et al.}(2007)]{Kim2007}
\reference
Kim, K. H., Y.-J. Moon, and K.-S. Cho (2007),
Prediction of the 1-AU arrival times of CME-associated
interplanetary shock: Evalution of an empirical interplanetary
shock propagation model, {\it J. Geophys. Res.}, {\it 112},
A05104, doi: 10.1029/2006JA011904, 2007.

  \bibitem[{\it Sheeley et al.}(1985a)]{Sheeley1985a}
\reference
Sheeley N. R., Jr., R. A. Howard, M. J. Koomen, D. J.
Michels, and R.  Schwenn (1985), Doppler scintillation
observations of interpanetary shocks within 0.3 AU, {\it J.
Geophys. Res.}, {\it 90}, 154.

  \bibitem[{\it Sheeley et al.}(1985b)]{Sheeley1985b}
\reference
Sheeley, N. R., Jr., R. A. Howard, M. J. Koomen, D. J.
Michels, H. V. Cane, S. W. Kahler, R. Schwenn. K. H. Muhlhauser,
and H. Rosenbauer,  Coronal mass ejections and interplanetary
shocks (1985), {\it J. Geophys. Res.}, {\it 90}, 163.

  \bibitem[{\it Trichtchenko et al.}(2007)]{Trichtchenko2007}
\reference
Trichtchenko, L., A. Zhukov, R. van der Linden, S. M.
Stankov, N. Jakowski, I. StanisJawska, G. Juchnikowski, P.
Wilkinson, G. Patterson, and A. W. P. Thomson (2007), November
2004 space weather events: Real-time observations and forecasts,
{\it Space Weather}, {\it 5}, S06001, doi:10.1029/2006SW000281.

  \bibitem[{\it Tsurutani et al.}(2008)]{Tsurutani2008}
\reference
Tsurutani, B. T., E. Echer, F. L. Guarnieri, and J. U.
Kozyra (2008), CAWSES November 7-8, 2004, superstorm: Complex
solar and interplanetary features in the post-solar maximum phase,
{\it Geophys. Res. Lett.}, {\it 35}, L06S05,
doi:10.1029/2007GL031473.

  \bibitem[{\it Wintoft et al.}(2005)]{Wintoft2005}
\reference
Wintoft P., M. Wik, H. Lundstedt, and L. Eliasson
(2005), Predictions of local ground geomagnetic field fluctuations
during the 7-10 November 2004 events studied with solar wind
driven models, {\it Ann. Geophys.}, {\it 23}, 3095-3101.

  \bibitem[{\it Yermolaev et al.}(2005)]{Yermolaev2005}
\reference Yermolaev, Yu. I., et al. (2005), A year later: Solar,
heliospheric and magnetospheric disturbances in November 2004,
{\it Geomagn. Aeron.}, {\it 45}(6), 681-719.

\end{thebibliography}
\end{document}